\documentclass{PoS}

\def\apj{ApJ}                 
\def\apjl{ApJ}                
\def\aap{A\&A}                

\def\mnras{MNRAS}             
\def\physrep{Phys.~Rep.}   

\usepackage{lineno}

\title{Search for very-high-energy photons from Gamma-ray bursts with HAWC} 

\ShortTitle{Follow up to GRBs}

\author{\speaker{Nissim Fraija} and M. Magdalena Gonz\'alez \\
The HAWC Collaboration\footnote{For collaboration list, see PoS(ICRC2019) 1177.}\\
{\itshape \href{https://www.hawc-observatory.org/collaboration/icrc2019.php}{https://www.hawc-observatory.org/collaboration/icrc2019.php}}\\
        Instituto de Astronom\'ia, UNAM.\\
        E-mail: \email{nifraija@astro.unam.mx}, \email{magda@astro.unam.mx}}

\abstract{Detections of very-high-energy (VHE; > 100 GeV) photons from Gamma-ray bursts (GRBs) can provide fundamental information on the involved radiative processes, physical composition of the ejecta and acceleration processes. The High Altitude Water Cherenkov (HAWC) gamma-ray observatory is the best gamma-ray instrument to study transient phenomena over a long period of time from 100 GeV to 100 TeV. 
Its large field of view and duty cycle (> 95\%) allow it to search blindly for sources of GRBs and to follow up on external alerts from satellite instruments, such as Fermi and Swift, searching for their VHE counterpart. We present results from the on-going GRB monitoring program and VHE upper limits of the latest interesting low-redshift and/or powerful bursts in HAWC's field of view: GRB 170206A, GRB 170817A, GRB 180720B and GRB 190114C.}

\FullConference{36th International Cosmic Ray Conference -ICRC2019-\\
		July 24th - August 1st, 2019\\
		Madison, WI, U.S.A.}

\begin{document}

\section{Introduction}
Gamma-ray bursts (GRBs) are the most luminous gamma-ray transient events in the Universe. They are non-repeating flashes usually associated with the core collapse of massive stars or the merger of compact object binaries when the duration of their prompt emission is longer or less than a few seconds, respectively \cite{2004IJMPA..19.2385Z, 2015PhR...561....1K}.  However, independently of the progenitor associated to the prompt emission, a long-lived afterglow emission is generated by the deceleration of the outflow in the circumburst medium (e.g., see \cite{2017ApJ...837..116B, 2019ApJ...872..118B,2019arXiv190405987B, 2019arXiv190706675F}).\\
The detection of very-high-energy (VHE; $\gtrsim $ 10 GeV) photons from GRBs and the record of their arrival times provide an important piece of information to estimate, for example, the baryonic composition of the outflow and the particle acceleration efficiency \cite{2015ApJ...804..105F, 2018ApJ...859...70F, 2016ApJ...818..190F,2016ApJ...831...22F, 2017ApJ...848...15F, 2017ApJ...848...94F, 2012ApJ...751...33F, 2012ApJ...755..127S, 2014MNRAS.437.2187F}. The Large Area Telescope (LAT) on-board the Fermi satellite has detected more than 150 GRBs which exhibit photons above $\geq$ 100 MeV but only $\sim$ 29 GRBs with VHE ($\geq 10$ GeV) photons \cite{Ajello_2019}. GRB 130427A presented photons with the highest energy ever detected, 73 GeV and 95 GeV, and  were observed 19 s and 244 s after the beginning of the burst, respectively \cite{2014Sci...343...42A}.  Additionally, a 52-GeV photon was related to GRB 160509A  \cite{2016GCN.19413....1L}. It arrived 77 s after the trigger. Recently, VHE photons with energies above of 300 GeV were detected from the long GRB 190114C \cite{2019ATel12390....1M}. The MAGIC telescopes observed GRB 190114C with a significance of $\sim$ 20$\sigma$ over the background for more than 1000 s \cite{2019ATel12390....1M}. Although VHE photons at TeV energies have been searched for by the Imaging Atmospheric Cherenkov Telescopes (IACTs),  only upper limits on the TeV flux have been derived  \cite{2007ApJ...667..358A, 2009ApJ...690.1068A}.\\
The High Altitude Water Cherenkov (HAWC) observatory followed-up GRB 130427A, one of the brightest and longest burst ever detected above 100 MeV. This burst took place when only 10\% of the final detector was operating and  under sub-optimal  observational conditions (see, \cite{2015ApJ...800...78A}).  Eight different time intervals during the prompt and long-lived emission were analyzed in order to searched for VHE emission, and although no statistically significant excess of counts was recorded in the GeV - TeV energy range, upper limits on the flux were placed \cite{2015ApJ...800...78A}.\\
In this work, we present the results from the on-going GRB monitoring program in HAWC and also the VHE upper limits of the latest interesting low-redshift and/or powerful bursts in the HAWC's field of view: GRB 170206A, GRB 170817A, GRB 180720B and GRB 190114C. In Section 2, we give a brief summary of the HAWC observatory and the on-going GRB monitoring program.  In section 3, we present the relevant information on GRB 170206A, GRB 170817A, GRB 180720B and GRB 190114C. In section 4, we show the results and a brief summary.

\section{GRB monitoring program}

The HAWC observatory, located at an altitude of 4,100 m a.s.l. in the volcano Sierra Negra in the state of Puebla, Mexico, is the best TeV gamma-ray observatory for studying transient phenomena at very-high energies, such as GRBs. Thanks to its wide field of view of $\sim$ 2 sr and its continuous operation (> 95$\%$ of duty cycle), HAWC constantly searches for TeV photons from GRBs detected by other instruments within its field of view. Since HAWC does not need to be re-pointed to the burst position, VHE data before, during, and after the burst trigger time are available, making it possible to test distinct model assumptions \cite{2017ApJ...841..100A}. It is worth noting that full temporal coverage only applies for bursts within the instantaneous field of view.\\
The on-going GRB monitoring program currently consists of two analyses: i) a rapid response follow-up of bursts notified by orbiting instruments and ii) a self-triggered all-sky search.   Both the triggered and self-triggered methods are applied to archival data in order to benefit from the latest reconstruction techniques.\\ 
The self-triggered all-sky searches permanently for VHE photons from bursts within three time windows: 0.2, 1 and 10 s. The analysis of rapid response follow-ups of notified bursts fixes the search window in accordance with the trigger time provided by the orbiting instrument. Details concerning the on-going GRB monitoring program can be found in \cite{2018arXiv180101437W}.  
\section{HAWC Upper Limits of interesting GRBs}
HAWC has not significantly detected emission from any GRB candidate to date. Therefore, upper limits in the GeV - TeV energy range for GRB 170206A, GRB 170817A, GRB 180720B and GRB 190114C have been set.  
Below we present the flux upper limits of the nearest/brightest bursts in the field of view of HAWC. The effect of  extragalactic background light (EBL) attenuation shown in \cite{2008A&A...487..837F} is used.
\subsection{GRB 170206A}
The Fermi GBM detected GRB 170206A at 10:51:57.70 UT on 2017 February 6th. The GBM light curve exhibited a short and bright burst with a duration of $T_{90}=1.2\,{\rm s}$ in the energy range of 50 - 300 keV \cite{2017GCN..20616....1V}. This burst was also detected at energies of MeV by Fermi LAT. The highest-energy photon reported was of 811 MeV and was observed 5.17 s after the trigger time \cite{2017GCN..20617....1D}.  This burst was also detected in gamma-rays by Konus-Wind. The Konus-Wind light curve displayed a multi-peaked structure with a high-energy emission seen up to ~15 MeV \cite{2017GCN..20625....1S}.  There is not measurement of redshift for this burst. \\
The HAWC upper limit of $2.82\times 10^{-6}\,{\rm erg\,cm^{-2}\,s^{-1}}$ at  6 s after the trigger time was calculated for $E_{\rm iso}=5.5\times10^{52}$ erg and an assumed z=0.3 \cite{2017ApJ...843...88A}, a likely value for a short burst. This burst was within the set of burst analyzed when searching for TeV emission coincident with the prompt emission. Therefore time windows of duration $T_{90}$, 3$\times T_{90}$ and 10$\times T_{90}$ were selected. The most restrictive upper limit is found and presented here for the time window of T90 \cite{2017ifs..confE..68D}. The isotropic energy  was computed taking into account the Band function parameters reported in the \cite{2017GCN.20616....1V}.

\subsection{GRB 170817A}
The gravitational-wave transient GW170817, associated with a binary neutron star system with a merger time of 12:41:04 UTC, 2017 August 17,  was detected by LIGO and Virgo experiments \cite{PhysRevLett.119.161101,2041-8205-848-2-L12}. Immediately,   GRB 170817A triggered on the GBM instrument at 12:41:06.47 UT on 2017 August 17 \cite{2017GCN.21520....1V}.  The Fermi GBM located a weak gamma-ray flux at R.A.$=176.8$ and DEC$=-39.8$ (J2000) with an uncertainty
of $11.6^\circ$. The gamma-ray light curve in the 50-300 keV energy range displayed a weak short pulse with a duration of $T_{90}=2\,{\rm s}$ \cite{2017GCN.21520....1V, 2017ApJ...848L..14G}. A large multi-wavelength campaign followed up this burst. In the X-ray bands, this burst was detected by the Chandra and XMM-Newton satellites \cite{troja2017a, 2018arXiv180103531M}. In optical bands, GRB 170817A was detected by the Hubble Space Telescope \cite{2018arXiv180102669L, 2018arXiv180103531M}. In the radio wavelengths, GRB 170817A was detected by Very Large Array \cite{2041-8205-848-2-L12}.  The host galaxy associated with this burst was located with a redshift of $z\simeq 0.01$ \cite{2017arXiv171005452C,2017ApJ...848L..20M}. A miscellaneous set of models based on external shocks  were developed to explain this burst \cite{2017arXiv171111573M, 2017arXiv171203237L, 2019ApJ...871..200F, 2019ApJ...871..123F, 2019arXiv190600502F, 2019arXiv190407732F}.\\

Unfortunately, this burst entered the HAWC's field of view $\sim$ 8 hours after the GW trigger \cite{2017ApJ...848L..12A}. Thus, the flux upper limit was set for the first transit of the burst location after the trigger time \cite{2017ApJ...848L..12A,Israel_ICRC}. Taking into account that the atypical multi-wavelength fluxes peaked at $\sim$ 120 days after the trigger time, a search for a TeV counterpart for much longer time periods than $T_{90}$ was performed, for details see \cite{2019ICRC...36..681G}. The most constraining flux upper limit found was $3.37\times 10^{-12}\,{\rm erg\,cm^{-2}\,s^{-1}}$  for an energy range of 7 - 170 TeV and the time period of 10 - 110 days after the merger time.

Von Kienlin et. al. \cite{2019arXiv190106158V} identified 13 GRBs  with similar features to GRB 170817A. Then, the search for TeV emission within days was extended to the 2 other bursts (out of the 13) in the HAWC's field of view, see \cite{2019ICRC...36..681G}.

\subsection{GRB 180720B}

GRB 180720B triggered on the Swift BAT at 14:21:44 UT on 2018 July 20 \cite{2018GCN.22973....1P}. This burst was located at R.A.$=00h\,02m\,07s$ and DEC$=-02d 56' 0"$ with an uncertainty of 3 arcmin \cite{2018GCN.22998....1B}. The BAT light curve exhibited a multi-peaked structure with a duration of $T_{90}= 150\,{\rm s}$ \cite{2018GCN.22973....1P}.  Promptly, GRB 180720B was followed up by X-ray Telescope (XRT) and the Ultra-Violet/Optical Telescope (UVOT), by Fermi GMB and LAT \cite{2018GCN.22981....1R, 2019arXiv190513572F}, by Konus-Wind \cite{2018GCN.23011....1F}, by CALET Gamma-ray Monitor \cite{2018GCN.23042....1C}  and several orbiting and ground instruments in optical and radio bands \cite{2018GCN.23040....1I}. The host galaxy of GRB 180720B was located to have a redshift of $z=0.654$ \cite{2018GCN.22996....1V}.

The VHE upper limit of $1.81\,\times 10^{-8}\,{\rm erg\,cm^{-2}\,s^{-1}}$ is reported  for an energy range of 2 - 60 TeV.  The observations of HAWC experiment began 18 hours after the trigger time. Thus, the flux upper limit is given for the first transit in the HAWC's field of view (lasted 4 hours).

\subsection{GRB 190114C}

The Swift BAT instrument triggered on GRB 190114C on January 14, 2019 at 20:57:03 UTC (trigger 883832) \cite{2019GCN.23688....1G}.  This burst was located at R.A.$=03h\,38m\,02s$ and DEC$=-26d 56' 18"$ (J2000) with an uncertainty of 3 arcmin. The BAT light curve displayed a very bright multi-peaked structure with a duration of $T_{90}=25\, {\rm s}$ \cite{2019GCN.23688....1G}. GRB 190114C was also detected by Fermi GBM \cite{2019GCN.23709....1D}, by Fermi LAT \cite{2019GCN.23709....1D, 2019ApJ...879L..26F}, by Swift XRT \cite{2019GCN.23688....1G, 2019GCN.23704....1O},  by Swift UVOT \cite{2019GCN.23688....1G, 2019GCN.23725....1S}, by INTEGRAL \cite{2019GCN.23714....1M},  by AGILE satellite \cite{2019GCN.23712....1U}, by Konus-Wind \cite{2019GCN.23737....1F} and by a massive campaign of optical instruments and telescopes   \cite{2019GCN.23690....1T}. For the first time an excess of gamma-ray events with a significance of $\>$20 $\sigma$ was detected during the first 20 minutes and photons with  energies above 300 GeV were reported by MAGIC collaboration from GRB 190114C  \cite{2019GCN.23701....1M}. The host galaxy of this burst was located at a redshift of  $z$ = 0.42 \cite{2019GCN.23692....1U,2019GCN.23695....1S}.

The VHE upper limit of $4.46\times 10^{-8}\,{\rm erg\,cm^{-2}\,s^{-1}}$  for an energy range of 7 - 170 TeV is reported. The observations of HAWC observatory lasted 2 hours and began 5 hours after the trigger time. Similarly to GRB 170206A and GRB 180720B, the flux upper limit is given for the first transit in the HAWC's field of view.

The summary of the VHE upper limits are reported in Table 1.

\begin{table}[h!]
\centering
\caption{VHE Upper limits calculated for some GRBs. The spectral power-law index of 2.5 was used in all cases. The integration times are given in column 3.}\label{upli}
\begin{tabular}{lccc}
\hline\hline
Bursts & Energy Range & Observation & Upper Limit\\
& & & [$10^{-8}\,{\rm erg\,cm^{-2}\,s^{-1}}$]\\
\hline
GRB 170206A & (80 - 800) GeV & 6 - 8 s & $2.82\times10^2$ \\ 
GRB 170817A & (7 - 170) TeV & 10 - 110 days  & $3.37\times 10^{-4}$ \\ 
GRB 180720B & (2 - 60) TeV & 18 - 22 hours & $1.81$ \\ 
GRB 190114C  & (7 - 170) TeV & 5 - 7 hours & $4.46$\\
\hline
\hline
\end{tabular}
\end{table}

\section{Summary}

We analyzed data taken  by the HAWC gamma ray observatory to search for VHE photons from GRB 170206A, GRB 170817A, GRB 180720B and GRB 190114C, all of which were inside HAWC's field of view and were reported by the orbiting instruments and ground telescopes.   These bursts occurred from February 2017 to January 2019.  Although no statistically significant excess of counts was detected by HAWC, VHE upper limits in the GeV - TeV energy were derived around the positions of GRB 170206A, GRB 170817A, GRB 180721A and GRB 190114C.

HAWC continues to monitor the whole sky in search of signals from potential burst candidates.  

\section{Acknowledgements}

We acknowledge the support from: the US National Science Foundation (NSF) the US Department of Energy Office of High-Energy Physics; 
the Laboratory Directed Research and Development (LDRD) program of Los Alamos National Laboratory; 
Consejo Nacional de Ciencia y Tecnolog\'{\i}a (CONACyT), M{\'e}xico (grants 271051, 232656, 260378, 179588, 239762, 254964, 271737, 258865, 243290, 132197, 281653)(C{\'a}tedras 873, 1563, 341), Laboratorio Nacional HAWC de rayos gamma; 
L'OREAL Fellowship for Women in Science 2014; 
Red HAWC, M{\'e}xico; 
DGAPA-UNAM (grants AG100317, IN111315, IN111716-3, IA102715, IN111419, IA102019, IN112218); 
VIEP-BUAP; 
PIFI 2012, 2013, PROFOCIE 2014, 2015; 
the University of Wisconsin Alumni Research Foundation; 
the Institute of Geophysics, Planetary Physics, and Signatures at Los Alamos National Laboratory; 
Polish Science Centre grant DEC-2014/13/B/ST9/945, DEC-2017/27/B/ST9/02272; 
Coordinaci{\'o}n de la Investigaci{\'o}n Cient\'{\i}fica de la Universidad Michoacana; Royal Society - Newton Advanced Fellowship 180385. Thanks to Scott Delay, Luciano D\'{\i}az and Eduardo Murrieta for technical support.


\begin{thebibliography}{10}

\bibitem{2004IJMPA..19.2385Z}
B.~{Zhang} and P.~{M{\'e}sz{\'a}ros}, \emph{{Gamma-Ray Bursts: progress,
  problems and prospects}},
  \href{https://doi.org/10.1142/S0217751X0401746X}{\emph{International Journal
  of Modern Physics A} {\bfseries 19} (2004) 2385}
  [\href{https://arxiv.org/abs/arXiv:astro-ph/0311321}{{\ttfamily
  arXiv:astro-ph/0311321}}].

\bibitem{2015PhR...561....1K}
P.~{Kumar} and B.~{Zhang}, \emph{{The physics of gamma-ray bursts and
  relativistic jets}},
  \href{https://doi.org/10.1016/j.physrep.2014.09.008}{\emph{\physrep}
  {\bfseries 561} (2015) 1} [\href{https://arxiv.org/abs/1410.0679}{{\ttfamily
  1410.0679}}].

\bibitem{2017ApJ...837..116B}
R.~L. {Becerra}, A.~M. {Watson}, W.~H. {Lee}, N.~{Fraija}, N.~R. {Butler},
  J.~S. {Bloom} et~al., \emph{{Photometric Observations of Supernova 2013cq
  Associated with GRB 130427A}},
  \href{https://doi.org/10.3847/1538-4357/aa610f}{\emph{\apj} {\bfseries 837}
  (2017) 116} [\href{https://arxiv.org/abs/1702.04762}{{\ttfamily
  1702.04762}}].

\bibitem{2019ApJ...872..118B}
R.~L. {Becerra}, A.~M. {Watson}, N.~{Fraija}, N.~R. {Butler}, W.~H. {Lee},
  E.~{Troja} et~al., \emph{{Late Central-engine Activity in GRB 180205A}},
  \href{https://doi.org/10.3847/1538-4357/ab0026}{\emph{\apj} {\bfseries 872}
  (2019) 118} [\href{https://arxiv.org/abs/1901.06051}{{\ttfamily
  1901.06051}}].

\bibitem{2019arXiv190405987B}
R.~L. {Becerra}, S.~{Dichiara}, A.~M. {Watson}, E.~{Troja}, N.~I. {Fraija},
  A.~{Klotz} et~al., \emph{{Reverse Shock Emission Revealed in Early Photometry
  in the Candidate Short GRB 180418A}}, {\emph{arXiv e-prints} (2019) }
  [\href{https://arxiv.org/abs/1904.05987}{{\ttfamily 1904.05987}}].

\bibitem{2019arXiv190706675F}
N.~{Fraija}, R.~{Barniol Duran}, S.~{Dichiara} and P.~{Beniamini},
  \emph{{Synchrotron self-Compton as a likely mechanism of photons beyond the
  synchrotron limit in GRB 190114C}}, {\emph{arXiv e-prints} (2019)
  arXiv:1907.06675} [\href{https://arxiv.org/abs/1907.06675}{{\ttfamily
  1907.06675}}].

\bibitem{2015ApJ...804..105F}
N.~{Fraija}, \emph{{GRB 110731A: Early Afterglow in Stellar Wind Powered By a
  Magnetized Outflow}},
  \href{https://doi.org/10.1088/0004-637X/804/2/105}{\emph{\apj} {\bfseries
  804} (2015) 105} [\href{https://arxiv.org/abs/1503.07449}{{\ttfamily
  1503.07449}}].

\bibitem{2018ApJ...859...70F}
N.~{Fraija} and P.~{Veres}, \emph{{The Origin of the Optical Flashes: The Case
  Study of GRB 080319B and GRB 130427A}},
  \href{https://doi.org/10.3847/1538-4357/aabd79}{\emph{\apj} {\bfseries 859}
  (2018) 70} [\href{https://arxiv.org/abs/1804.02449}{{\ttfamily 1804.02449}}].

\bibitem{2016ApJ...818..190F}
N.~{Fraija}, W.~{Lee} and P.~{Veres}, \emph{{Modeling the Early Multiwavelength
  Emission in GRB130427A}},
  \href{https://doi.org/10.3847/0004-637X/818/2/190}{\emph{\apj} {\bfseries
  818} (2016) 190} [\href{https://arxiv.org/abs/1601.01264}{{\ttfamily
  1601.01264}}].

\bibitem{2016ApJ...831...22F}
N.~{Fraija}, W.~H. {Lee}, P.~{Veres} and R.~{Barniol Duran}, \emph{{Modeling
  the Early Afterglow in the Short and Hard GRB 090510}},
  \href{https://doi.org/10.3847/0004-637X/831/1/22}{\emph{\apj} {\bfseries 831}
  (2016) 22}.

\bibitem{2017ApJ...848...15F}
N.~{Fraija}, P.~{Veres}, B.~B. {Zhang}, R.~{Barniol Duran}, R.~L. {Becerra},
  B.~{Zhang} et~al., \emph{{Theoretical Description of GRB 160625B with
  Wind-to-ISM Transition and Implications for a Magnetized Outflow}},
  \href{https://doi.org/10.3847/1538-4357/aa8a72}{\emph{\apj} {\bfseries 848}
  (2017) 15} [\href{https://arxiv.org/abs/1705.09311}{{\ttfamily 1705.09311}}].

\bibitem{2017ApJ...848...94F}
N.~{Fraija}, W.~H. {Lee}, M.~{Araya}, P.~{Veres}, R.~{Barniol Duran} and
  S.~{Guiriec}, \emph{{Modeling the High-energy Emission in GRB 110721A and
  Implications on the Early Multiwavelength and Polarimetric Observations}},
  \href{https://doi.org/10.3847/1538-4357/aa8d65}{\emph{\apj} {\bfseries 848}
  (2017) 94} [\href{https://arxiv.org/abs/1709.06263}{{\ttfamily 1709.06263}}].

\bibitem{2012ApJ...751...33F}
N.~{Fraija}, M.~M. {Gonz{\'a}lez} and W.~H. {Lee}, \emph{{Synchrotron
  Self-Compton Emission as the Origin of the Gamma-Ray Afterglow Observed in
  GRB 980923}}, \href{https://doi.org/10.1088/0004-637X/751/1/33}{\emph{\apj}
  {\bfseries 751} (2012) 33} [\href{https://arxiv.org/abs/1201.3689}{{\ttfamily
  1201.3689}}].

\bibitem{2012ApJ...755..127S}
J.~R. {Sacahui}, N.~{Fraija}, M.~M. {Gonz{\'a}lez} and W.~H. {Lee}, \emph{{The
  Long and the Short of the High-energy Emission in GRB090926A: An External
  Shock}}, \href{https://doi.org/10.1088/0004-637X/755/2/127}{\emph{\apj}
  {\bfseries 755} (2012) 127}
  [\href{https://arxiv.org/abs/1203.1577}{{\ttfamily 1203.1577}}].

\bibitem{2014MNRAS.437.2187F}
N.~{Fraija}, \emph{{GeV-PeV neutrino production and oscillation in hidden jets
  from gamma-ray bursts}},
  \href{https://doi.org/10.1093/mnras/stt2036}{\emph{\mnras} {\bfseries 437}
  (2014) 2187} [\href{https://arxiv.org/abs/1310.7061}{{\ttfamily 1310.7061}}].

\bibitem{Ajello_2019}
M.~Ajello, M.~Arimoto, M.~Axelsson, L.~Baldini, G.~Barbiellini, D.~Bastieri
  et~al., \emph{A decade of gamma-ray bursts observed by fermi-{LAT}: The
  second {GRB} catalog},
  \href{https://doi.org/10.3847/1538-4357/ab1d4e}{\emph{The Astrophysical
  Journal} {\bfseries 878} (2019) 52}.

\bibitem{2014Sci...343...42A}
M.~{Ackermann}, M.~{Ajello}, K.~{Asano}, W.~B. {Atwood}, M.~{Axelsson},
  L.~{Baldini} et~al., \emph{{Fermi-LAT Observations of the Gamma-Ray Burst GRB
  130427A}}, \href{https://doi.org/10.1126/science.1242353}{\emph{Science}
  {\bfseries 343} (2014) 42}.

\bibitem{2016GCN.19413....1L}
F.~{Longo}, E.~{Bissaldi}, G.~{Vianello}, E.~{Moretti}, N.~{Omodei},
  J.~{Bregeon} et~al., \emph{{GRB 160509A: Fermi-LAT refined analysis.}},
  {\emph{GRB Coordinates Network, Circular Service, No.~19413, \#1 (2016)}
  {\bfseries 19413} (2016) }.

\bibitem{2019ATel12390....1M}
R.~{Mirzoyan}, \emph{{First time detection of a GRB at sub-TeV energies; MAGIC
  detects the GRB 190114C}}, {\emph{The Astronomer's Telegram} {\bfseries
  12390} (2019) }.

\bibitem{2007ApJ...667..358A}
J.~{Albert}, E.~{Aliu}, H.~{Anderhub}, P.~{Antoranz}, A.~{Armada},
  C.~{Baixeras} et~al., \emph{{MAGIC Upper Limits on the Very High Energy
  Emission from Gamma-Ray Bursts}},
  \href{https://doi.org/10.1086/520761}{\emph{\apj} {\bfseries 667} (2007) 358}
  [\href{https://arxiv.org/abs/astro-ph/0612548}{{\ttfamily
  astro-ph/0612548}}].

\bibitem{2009ApJ...690.1068A}
F.~{Aharonian}, A.~G. {Akhperjanian}, U.~{Barres DeAlmeida}, A.~R.
  {Bazer-Bachi}, B.~{Behera}, M.~{Beilicke} et~al., \emph{{HESS Observations of
  the Prompt and Afterglow Phases of GRB 060602B}},
  \href{https://doi.org/10.1088/0004-637X/690/2/1068}{\emph{\apj} {\bfseries
  690} (2009) 1068} [\href{https://arxiv.org/abs/0809.2334}{{\ttfamily
  0809.2334}}].

\bibitem{2015ApJ...800...78A}
A.~U. {Abeysekara}, R.~{Alfaro}, C.~{Alvarez}, J.~D. {{\'A}lvarez}, R.~{Arceo},
  J.~C. {Arteaga-Vel{\'a}zquez} et~al., \emph{{Search for Gamma-Rays from the
  Unusually Bright GRB 130427A with the HAWC Gamma-Ray Observatory}},
  \href{https://doi.org/10.1088/0004-637X/800/2/78}{\emph{\apj} {\bfseries 800}
  (2015) 78} [\href{https://arxiv.org/abs/1410.1536}{{\ttfamily 1410.1536}}].

\bibitem{2017ApJ...841..100A}
A.~U. {Abeysekara}, A.~{Albert}, R.~{Alfaro}, C.~{Alvarez}, J.~D.
  {{\'A}lvarez}, R.~{Arceo} et~al., \emph{{Daily Monitoring of TeV Gamma-Ray
  Emission from Mrk 421, Mrk 501, and the Crab Nebula with HAWC}},
  \href{https://doi.org/10.3847/1538-4357/aa729e}{\emph{\apj} {\bfseries 841}
  (2017) 100} [\href{https://arxiv.org/abs/1703.06968}{{\ttfamily
  1703.06968}}].

\bibitem{2018arXiv180101437W}
J.~{Wood}, \emph{{Results from the first one and a half years of the HAWC GRB
  program}}, {\emph{arXiv e-prints} (2018) arXiv:1801.01437}
  [\href{https://arxiv.org/abs/1801.01437}{{\ttfamily 1801.01437}}].

\bibitem{2008A&A...487..837F}
A.~{Franceschini}, G.~{Rodighiero} and M.~{Vaccari}, \emph{{Extragalactic
  optical-infrared background radiation, its time evolution and the cosmic
  photon-photon opacity}},
  \href{https://doi.org/10.1051/0004-6361:200809691}{\emph{\aap} {\bfseries
  487} (2008) 837} [\href{https://arxiv.org/abs/0805.1841}{{\ttfamily
  0805.1841}}].

\bibitem{2017GCN..20616....1V}
A.~{Von Kienlin} and O.~J. {Robert}, \emph{{GRB 170206A: Fermi GBM
  observation}}, {\emph{GRB Coordinates Network} {\bfseries 20616} (2017) 1}.

\bibitem{2017GCN..20617....1D}
F.~F. {Dirirsa} and {et al.}, \emph{{GRB 170206A: Fermi-LAT detection}},
  {\emph{GRB Coordinates Network} {\bfseries 20617} (2017) 1}.

\bibitem{2017GCN..20625....1S}
D.~{Svinkin} and {et al.}, \emph{{Konus-Wind observation of GRB 170206A}},
  {\emph{GRB Coordinates Network} {\bfseries 20625} (2017) 1}.

\bibitem{2017ApJ...843...88A}
R.~{Alfaro}, C.~{Alvarez}, J.~D. {{\'A}lvarez}, R.~{Arceo}, J.~C.
  {Arteaga-Vel{\'a}zquez} and {HAWC Collaboration}, \emph{{Search for
  Very-high-energy Emission from Gamma-Ray Bursts Using the First 18 Months of
  Data from the HAWC Gamma-Ray Observatory}},
  \href{https://doi.org/10.3847/1538-4357/aa756f}{\emph{\apj} {\bfseries 843}
  (2017) 88} [\href{https://arxiv.org/abs/1705.01551}{{\ttfamily 1705.01551}}].

\bibitem{2017ifs..confE..68D}
S.~{Dichiara}, M.~M. {Gonzalez}, N.~{Fraija} and {HAWC Collaboration},
  \emph{{Constraints on microphysical parameters of GRBs using HAWC}},  in
  \emph{Proceedings of the 7th International Fermi Symposium}, p.~68, Oct,
  2017.

\bibitem{2017GCN.20616....1V}
A.~{von Kienlin} and O.~J. {Roberts}, \emph{{GRB 170206A: Fermi GBM
  observation.}}, {\emph{GRB Coordinates Network, Circular Service, No.~20616,
  \#1 (2017)} {\bfseries 20616} (2017) }.

\bibitem{PhysRevLett.119.161101}
{\scshape LIGO Scientific Collaboration and Virgo Collaboration} collaboration,
  \emph{Gw170817: Observation of gravitational waves from a binary neutron star
  inspiral}, \href{https://doi.org/10.1103/PhysRevLett.119.161101}{\emph{Phys.
  Rev. Lett.} {\bfseries 119} (2017) 161101}.

\bibitem{2041-8205-848-2-L12}
B.~P. Abbott, R.~Abbott, T.~D. Abbott and et~al., \emph{Multi-messenger
  observations of a binary neutron star merger}, {\emph{The Astrophysical
  Journal Letters} {\bfseries 848} (2017) L12}.

\bibitem{2017GCN.21520....1V}
A.~{von Kienlin}, C.~{Meegan} and A.~{Goldstein}, \emph{{GRB 170817A: Fermi GBM
  detection.}}, {\emph{GRB Coordinates Network, Circular Service, No.~21520,
  \#1 (2017)} {\bfseries 21520} (2017) }.

\bibitem{2017ApJ...848L..14G}
A.~{Goldstein}, P.~{Veres}, E.~{Burns}, M.~S. {Briggs}, R.~{Hamburg},
  D.~{Kocevski} et~al., \emph{{An Ordinary Short Gamma-Ray Burst with
  Extraordinary Implications: Fermi-GBM Detection of GRB 170817A}},
  \href{https://doi.org/10.3847/2041-8213/aa8f41}{\emph{\apjl} {\bfseries 848}
  (2017) L14} [\href{https://arxiv.org/abs/1710.05446}{{\ttfamily
  1710.05446}}].

\bibitem{troja2017a}
E.~Troja, L.~Piro, H.~van Eerten and et~al., \emph{The x-ray counterpart to the
  gravitational-wave event gw170817},
  \href{https://doi.org/10.1038/nature24290}{\emph{Nature} {\bfseries 000}
  (2017) 1}.

\bibitem{2018arXiv180103531M}
R.~{Margutti}, K.~D. {Alexander}, X.~{Xie}, L.~{Sironi}, B.~D. {Metzger},
  A.~{Kathirgamaraju} et~al., \emph{{The Binary Neutron Star event LIGO/VIRGO
  GW170817 a hundred days after merger: synchrotron emission across the
  electromagnetic spectrum}}, {\emph{ArXiv e-prints} (2018) }
  [\href{https://arxiv.org/abs/1801.03531}{{\ttfamily 1801.03531}}].

\bibitem{2018arXiv180102669L}
J.~D. {Lyman}, G.~P. {Lamb}, A.~J. {Levan}, I.~{Mandel}, N.~R. {Tanvir},
  S.~{Kobayashi} et~al., \emph{{The optical afterglow of the short gamma-ray
  burst associated with GW170817}}, {\emph{ArXiv e-prints} (2018) }
  [\href{https://arxiv.org/abs/1801.02669}{{\ttfamily 1801.02669}}].

\bibitem{2017arXiv171005452C}
D.~A. {Coulter}, R.~J. {Foley}, C.~D. {Kilpatrick}, M.~R. {Drout}, A.~L.
  {Piro}, B.~J. {Shappee} et~al., \emph{{Swope Supernova Survey 2017a (SSS17a),
  the Optical Counterpart to a Gravitational Wave Source}}, {\emph{ArXiv
  e-prints} (2017) } [\href{https://arxiv.org/abs/1710.05452}{{\ttfamily
  1710.05452}}].

\bibitem{2017ApJ...848L..20M}
R.~{Margutti}, E.~{Berger}, W.~{Fong}, C.~{Guidorzi}, K.~D. {Alexander}, B.~D.
  {Metzger} et~al., \emph{{The Electromagnetic Counterpart of the Binary
  Neutron Star Merger LIGO/Virgo GW170817. V. Rising X-Ray Emission from an
  Off-axis Jet}}, \href{https://doi.org/10.3847/2041-8213/aa9057}{\emph{\apjl}
  {\bfseries 848} (2017) L20}
  [\href{https://arxiv.org/abs/1710.05431}{{\ttfamily 1710.05431}}].

\bibitem{2017arXiv171111573M}
K.~P. {Mooley}, E.~{Nakar}, K.~{Hotokezaka}, G.~{Hallinan}, A.~{Corsi}, D.~A.
  {Frail} et~al., \emph{{A mildly relativistic wide-angle outflow in the
  neutron star merger GW170817}}, {\emph{ArXiv e-prints} (2017) }
  [\href{https://arxiv.org/abs/1711.11573}{{\ttfamily 1711.11573}}].

\bibitem{2017arXiv171203237L}
D.~{Lazzati}, R.~{Perna}, B.~J. {Morsony}, D.~{L{\'o}pez-C{\'a}mara},
  M.~{Cantiello}, R.~{Ciolfi} et~al., \emph{{Late time afterglow observations
  reveal a collimated relativistic jet in the ejecta of the binary neutron star
  merger GW170817}}, {\emph{ArXiv e-prints} (2017) }
  [\href{https://arxiv.org/abs/1712.03237}{{\ttfamily 1712.03237}}].

\bibitem{2019ApJ...871..200F}
N.~{Fraija}, A.~C.~C.~d.~E.~S. {Pedreira} and P.~{Veres}, \emph{{Light Curves
  of a Shock-breakout Material and a Relativistic Off-axis Jet from a Binary
  Neutron Star System}},
  \href{https://doi.org/10.3847/1538-4357/aaf80e}{\emph{\apj} {\bfseries 871}
  (2019) 200}.

\bibitem{2019ApJ...871..123F}
N.~{Fraija}, F.~{De Colle}, P.~{Veres}, S.~{Dichiara}, R.~{Barniol Duran},
  A.~{Galvan-Gamez} et~al., \emph{{The Short GRB 170817A: Modeling the Off-axis
  Emission and Implications on the Ejecta Magnetization}},
  \href{https://doi.org/10.3847/1538-4357/aaf564}{\emph{\apj} {\bfseries 871}
  (2019) 123}.

\bibitem{2019arXiv190600502F}
N.~{Fraija}, F.~{De Colle}, P.~{Veres}, S.~{Dichiara}, R.~{Barniol Duran},
  A.~C. C. d. E.~S. {Pedreira} et~al., \emph{{Description of atypical bursts
  seen slightly off-axis}}, {\emph{arXiv e-prints} (2019) arXiv:1906.00502}
  [\href{https://arxiv.org/abs/1906.00502}{{\ttfamily 1906.00502}}].

\bibitem{2019arXiv190407732F}
N.~{Fraija}, D.~{Lopez-Camara}, A.~C. C. d. E.~S. {Pedreira}, B.~{Betancourt
  Kamenetskaia}, P.~{Veres} and S.~{Dichiara}, \emph{{Signatures from a Cocoon
  and an off-axis material ejected in a merger of compact objects: An
  analytical approach}}, {\emph{arXiv e-prints} (2019) arXiv:1904.07732}
  [\href{https://arxiv.org/abs/1904.07732}{{\ttfamily 1904.07732}}].

\bibitem{2017ApJ...848L..12A}
B.~P. {Abbott}, R.~{Abbott}, T.~D. {Abbott}, F.~{Acernese}, K.~{Ackley},
  C.~{Adams} et~al., \emph{{Multi-messenger Observations of a Binary Neutron
  Star Merger}}, \href{https://doi.org/10.3847/2041-8213/aa91c9}{\emph{\apjl}
  {\bfseries 848} (2017) L12}
  [\href{https://arxiv.org/abs/1710.05833}{{\ttfamily 1710.05833}}].

\bibitem{Israel_ICRC}
I.~{Martinez-Castellanos} and {for the HAWC Collaboration}, \emph{{Search for
  very-high-energy gamma-ray counterparts of gravitational waves with HAWC}}, .

\bibitem{2019ICRC...36..681G}
A.~{Galv{\'a}n}, N.~{Fraija} and M.~M. {Gonz{\'a}lez}, \emph{{Search for
  very-high-energy emission with HAWC from GW170817 event}},  in \emph{36th
  International Cosmic Ray Conference (ICRC2019)}, vol.~36 of
  \emph{International Cosmic Ray Conference}, p.~681, Jul, 2019.

\bibitem{2019arXiv190106158V}
A.~{von Kienlin}, P.~{Veres}, O.~J. {Roberts}, R.~{Hamburg}, E.~{Bissaldi},
  M.~S. {Briggs} et~al., \emph{{Fermi GBM GRBs with characteristics similar to
  GRB 170817A}}, {\emph{arXiv e-prints} (2019) arXiv:1901.06158}
  [\href{https://arxiv.org/abs/1901.06158}{{\ttfamily 1901.06158}}].

\bibitem{2018GCN.22973....1P}
D.~{Palmer}, M.~H. {Siegel}, D.~N. {Burrows} and {et al.}., \emph{{GRB 180720B:
  Swift detection of a burst}}, {\emph{GRB Coordinates Network, Circular
  Service, No.~22973, \#1 (2018)} {\bfseries 22973} (2018) }.

\bibitem{2018GCN.22998....1B}
S.~D. {Barthelmy}, J.~R. {Cummings}, H.~A. {Krimm}, A.~Y. {Lien}, C.~B.
  {Markwardt}, D.~M. {Palmer} et~al., \emph{{GRB 180720B: Swift-BAT refined
  analysis.}}, {\emph{GRB Coordinates Network, Circular Service, No.~22998, \#1
  (2018)} {\bfseries 22998} (2018) }.

\bibitem{2018GCN.22981....1R}
O.~J. {Roberts} and C.~{Meegan}, \emph{{GRB 180720B: Fermi GBM observation.}},
  {\emph{GRB Coordinates Network, Circular Service, No.~22981, \#1 (2018)}
  {\bfseries 22981} (2018) }.

\bibitem{2019arXiv190513572F}
N.~{Fraija}, S.~{Dichiara}, A.~C. C. d. E.~S. {Pedreira}, A.~{Galvan-Gamez},
  R.~L. {Becerra}, A.~{Montalvo} et~al., \emph{{Modeling observations of GRB
  180720B: From radio to GeV gamma-rays}}, {\emph{arXiv e-prints} (2019)
  arXiv:1905.13572} [\href{https://arxiv.org/abs/1905.13572}{{\ttfamily
  1905.13572}}].

\bibitem{2018GCN.23011....1F}
D.~{Frederiks}, S.~{Golenetskii}, R.~{Aptekar}, A.~{Kozlova}, A.~{Lysenko},
  D.~{Svinkin} et~al., \emph{{Konus-Wind observation of GRB 180720B.}},
  {\emph{GRB Coordinates Network, Circular Service, No.~23011, \#1 (2018)}
  {\bfseries 23011} (2018) }.

\bibitem{2018GCN.23042....1C}
M.~L. {Cherry}, A.~{Yoshida}, T.~{Sakamoto}, S.~{Sugita}, Y.~{Kawakubo},
  A.~{Tezuka} et~al., \emph{{GRB 180720B: CALET Gamma-Ray Burst Monitor
  detection.}}, {\emph{GRB Coordinates Network, Circular Service, No.~23042,
  \#1 (2018)} {\bfseries 23042} (2018) }.

\bibitem{2018GCN.23040....1I}
L.~{Izzo}, D.~A. {Kann}, A.~{de Ugarte Postigo}, C.~C. {Thoene}, K.~{Bensch},
  M.~{Blazek} et~al., \emph{{GRB 180720B: OAJ optical observations.}},
  {\emph{GRB Coordinates Network, Circular Service, No.~23040, \#1 (2018)}
  {\bfseries 23040} (2018) }.

\bibitem{2018GCN.22996....1V}
P.~M. {Vreeswijk}, D.~A. {Kann}, K.~E. {Heintz}, A.~{de Ugarte Postigo},
  B.~{Milvang-Jensen}, D.~B. {Malesani} et~al., \emph{{GRB 180720B:
  VLT/X-shooter redshift.}}, {\emph{GRB Coordinates Network, Circular Service,
  No.~22996, \#1 (2018)} {\bfseries 22996} (2018) }.

\bibitem{2019GCN.23688....1G}
J.~D. e.~a. {Gropp}, \emph{{GRB 190114C: }}, {\emph{GRB Coordinates Network,
  Circular Service, No.~23688} {\bfseries 23688} (2019) }.

\bibitem{2019GCN.23709....1D}
D.~e.~a. {Kocevski}, \emph{{GRB 190114C: }}, {\emph{GRB Coordinates Network,
  Circular Service, No.~23709} {\bfseries 23709} (2019) }.

\bibitem{2019ApJ...879L..26F}
N.~{Fraija}, S.~{Dichiara}, A.~C. C. d. E.~S. {Pedreira}, A.~{Galvan-Gamez},
  R.~L. {Becerra}, R.~{Barniol Duran} et~al., \emph{{Analysis and Modeling of
  the Multi-wavelength Observations of the Luminous GRB 190114C}},
  \href{https://doi.org/10.3847/2041-8213/ab2ae4}{\emph{\apjl} {\bfseries 879}
  (2019) L26} [\href{https://arxiv.org/abs/1904.06976}{{\ttfamily
  1904.06976}}].

\bibitem{2019GCN.23704....1O}
J.~P. e.~a. {Osborne}, \emph{{GRB 190114C: }}, {\emph{GRB Coordinates Network,
  Circular Service, No.~23704} {\bfseries 23704} (2019) }.

\bibitem{2019GCN.23725....1S}
M.~H. e.~a. {Siegel}, \emph{{GRB 190114C: }}, {\emph{GRB Coordinates Network,
  Circular Service, No.~23725} {\bfseries 23725} (2019) }.

\bibitem{2019GCN.23714....1M}
P.~{Minaev} and A.~{Pozanenko}, \emph{{GRB 190114C: SPI-ACS/INTEGRAL extended
  emission detection.}}, {\emph{GRB Coordinates Network, Circular Service,
  No.~23714, \#1 (2019)} {\bfseries 23714} (2019) }.

\bibitem{2019GCN.23712....1U}
A.~{Ursi}, M.~{Tavani}, M.~{Marisaldi}, N.~{Parmiggiani}, F.~{Longo},
  A.~{Argan} et~al., \emph{{GRB 190114C: AGILE/MCAL detection.}}, {\emph{GRB
  Coordinates Network, Circular Service, No.~23712, \#1 (2019)} {\bfseries
  23712} (2019) }.

\bibitem{2019GCN.23737....1F}
D.~{Frederiks}, S.~{Golenetskii}, R.~{Aptekar}, A.~{Kozlova}, A.~{Lysenko},
  D.~{Svinkin} et~al., \emph{{Konus-Wind observation of GRB 190114C.}},
  {\emph{GRB Coordinates Network, Circular Service, No.~23737, \#1 (2019)}
  {\bfseries 23737} (2019) }.

\bibitem{2019GCN.23690....1T}
N.~e.~a. {Tyurina}, \emph{{GRB 190114C: }}, {\emph{GRB Coordinates Network,
  Circular Service, No.~23690} {\bfseries 23690} (2019) }.

\bibitem{2019GCN.23701....1M}
R.~e.~a. {Mirzoyan}, \emph{{GRB 190114C: }}, {\emph{GRB Coordinates Network,
  Circular Service, No.~23701} {\bfseries 23701} (2019) }.

\bibitem{2019GCN.23692....1U}
A.~e.~a. {Ugarte Postigo}, \emph{{GRB 190114C: }}, {\emph{GRB Coordinates
  Network, Circular Service, No.~23692} {\bfseries 23692} (2019) }.

\bibitem{2019GCN.23695....1S}
J.~e.~a. {Selsing }, \emph{{GRB 190114C: }}, {\emph{GRB Coordinates Network,
  Circular Service, No.~23695} {\bfseries 23695} (2019) }.

\end{thebibliography}

\providecommand{\href}[2]{#2}\begingroup\raggedright\endgroup

\end{document}